# Conscious Access as Continuous-to-Discrete Translation

Tianming Yang

Institute of Neuroscience, State Key Laboratory of Brain Cognition and Brain-inspired Intelligence Technology, Center for Excellence in Brain Science and Intelligence Technology, Chinese Academy of Sciences

**Correspondence to:**

Tianming Yang, Ph.D.

tyang@ion.ac.cn
Institute of Neuroscience, Center for Excellence in Brain Science and Intelligence Technology
Chinese Academy of Sciences
System Neuroscience Building, Rm 302
320 Yueyang Rd
Shanghai
China

# Abstract

The scientific study of consciousness frequently stalls on ontological debates regarding the "Hard Problem." This paper proposes a pragmatic pivot. Rather than asking what consciousness is metaphysically, we ask how modeling conscious access as a specific computational transformation may address existing bottlenecks in neuroscience and artificial intelligence. We introduce the Continuous/Discrete (C/D) framework, which holds that the brain implements two distinct processing regimes: System C, a distributed sensory-motor network operating over continuous, high-dimensional manifolds, and System D, a centralized engine structured around discrete, scale-invariant symbols. We argue that conscious access requires a structure-preserving translation between these regimes, which maps localized continuous states onto discrete symbolic tokens, coupled with an inverse projection that grounds those tokens back into sensorimotor dynamics. By formalizing conscious access as this continuous-to-discrete conversion, we derive a unified set of testable predictions centered on representational geometry, specifically, on a measurable collapse from graded similarity structures to low-dimensional categorical equivalence classes. These predictions explicitly differentiate our account from Global Neuronal Workspace Theory, Integrated Information Theory, Predictive Processing, and Higher-Order Theories, shifting the focus from ontological status to computational mechanism. Beyond neuroscience, the framework provides a principled architecture for neuro-symbolic artificial intelligence. We argue that treating conscious access as translational computation offers a pragmatic, empirically tractable pathway forward, clarifying what conscious states functionally accomplish without requiring resolution of the hard problem of phenomenology.

# 1. Introduction: The Case for Pragmatism in Consciousness Research

## 1.1 The Ontological Gridlock

More than three decades after Chalmers (1995) articulated the "Hard Problem," the scientific study of consciousness remains conspicuously entangled with the metaphysical question it was meant to set aside. Empirical programs that began as tractable investigations of neural correlates have, over time, accreted ontological commitments that now shape the experiments researchers are willing to run. Global Neuronal Workspace Theory (GNWT), for instance, began as a theory of information broadcast across frontoparietal networks (Dehaene & Naccache, 2001), but has increasingly been asked to adjudicate whether global availability is sufficient for subjective experience, or merely necessary. Higher-Order Theories (HOT), originally accounts of the functional role of metarepresentation (Rosenthal, 2005), have similarly been pressed to specify whether prefrontal re-representation is sufficient for phenomenal experience or merely enables report (Lau & Rosenthal, 2011; Brown, Lau, & LeDoux, 2019). Integrated Information Theory (IIT) argues that $\Phi$, a measure of information integration, is not merely a correlate of consciousness but, on the strong reading, is identical to it (Tononi et al., 2016). Predictive Processing (PP) frameworks, while ostensibly neutral on phenomenality, have been extended into claims that the precision-weighting of prediction errors constitutes the "felt quality" of experience (Clark, 2019; Seth, 2021).

A recent adversarial collaboration between GNWT and IIT proponents (Cogitate Consortium et al., 2025) produced data that proponents of both sides could, and did, interpret as partial vindication. This is not a failure of experimental design. It is a structural feature of theories whose core predictions concern the metaphysical status of neural states rather than their computational role. When theories disagree about what consciousness is, rather than what it does, experimental dissociations rarely settle the question.

We suggest that this gridlock follows from a widely shared, and rarely examined, assumption: that a theory of consciousness must, to be complete, specify the conditions under which a physical system gives rise to subjective experience. On this view, any framework that merely characterizes the computational profile of conscious states is explanatorily incomplete. We think this assumption has become a methodological liability. It conflates two problems, the explanation of phenomenality and the characterization of cognitive function, that are separable and that progress at very different rates.

## 1.2 The Pragmatic Pivot

Historically, this pragmatic approach has proven useful. Newton famously declined to speculate on the mechanism of gravitational attraction, treating gravity as a mathematically specified relation between masses whose deeper nature could be left to later inquiry. Similarly, early molecular biology made decisive progress by treating the gene as a functional unit of heredity long before its physical substrate was identified.

We propose an analogous move for consciousness science. Rather than asking what consciousness is, we ask: what does treating conscious access as a specific computational transformation allow us to explain, predict, and build? This is the methodological stance of scientific pragmatism in the tradition of Peirce, James, and, more recently, Putnam (Putnam, 1995; Misak, 2013). Our aim is not to resolve metaphysical puzzles but to generate testable predictions and unify previously disparate findings.

This pivot does not mean that questions about phenomenal experience are meaningless. Instead, we are claiming something more tractable. If we can specify what transformation consciousness performs, what its inputs are, what its outputs are, and what representational format mediates between them, we will make significant progress in the scientific understanding of consciousness and may deliver traction in adjacent fields, notably artificial intelligence, where ontological debates have so far produced little.

## 1.3 The Core Thesis: Conscious Access as Representational Translation

The pragmatic pivot demands that we look past phenomenological labels, such as "awareness" or "feeling", to identify the specific computational structure of conscious access. If consciousness is a mechanism, it must be supported by a specific type of data processing. To uncover this architecture, we must first diagnose why existing frameworks fail to identify the neural "joint" at which computation shifts.

The central proposal of this paper is that much of the computational utility of conscious access can be understood in terms of a specific kind of data transformation. The brain confronts what we will argue is a fundamental representational mismatch. The world it must perceive and act upon is presented through analog, continuous, high-dimensional, and context-dependent signals arriving via sensory organs and issued through motor effectors. Yet many of the cognitive operations that distinguish human intelligence, including planning, counterfactual reasoning, language, and logic, appear to require discrete, compositional, scale-invariant symbols that can be manipulated without reference to their sensory origins. Modeling conscious access as the mechanism by which the brain bridges this gap offers, we suggest, a pragmatic approach to advancing our understanding of consciousness. It brackets the metaphysical questions surrounding phenomenal experience in favor of a tractable computational characterization.

More specifically, we propose that the brain operates with two distinct representational formats, which we term System C (Continuous) and System D (Discrete) (**Table 1**). System C is the distributed, massively parallel sensory-motor engine that encodes the world as high-dimensional continuous manifolds, tightly coupled to spatiotemporal structure. System D is a centralized, serial, symbolic engine whose representations are discrete categories that have been explicitly decoupled from the continuous substrate from which they were extracted. Conscious access is neither a property of System C nor of System D in isolation. It requires the structural translation between them (**Figure 1**).

This framing has three immediate virtues. First, it is computationally explicit. The C-to-D translation is a specific operation with identifiable signatures in neural representational geometry, accessible to analytic methods. Second, it generates predictions that distinguish it from GNWT,

IIT, PP, and HOT as competing computational hypotheses. Third, the C-to-D translation in biological brains is also a central unsolved neuro-symbolic integration problem in artificial intelligence. The framework also has significant implications for understanding psychiatric conditions characterized by breakdowns in symbolic abstraction or its grounding in sensorimotor experience, an equally consequential application that we reserve for a dedicated future treatment.

The remainder of this paper develops this proposal in detail. **Section 2** diagnoses the representational bottleneck that motivates the C/D distinction and formalizes the C/D framework, specifying the computational properties of each system. **Section 3** presents the proposal that conscious access can be modeled as the computational operation that performs the information transfer between the two systems. **Section 4** derives empirical predictions and contrasts them explicitly with those of GNWT, IIT, PP, and HOT. **Section 5** applies the framework to the design of neuro-symbolic AGI.

Our aim throughout is not to declare the Hard Problem solved, dissolved, or irrelevant. It is to demonstrate that a well-specified computational model of conscious access can do substantial scientific work.

| System | C (Continuous) | D (Discrete) |
|---|---|---|
| **Representational format** | Continuous, high-dimensional neural manifolds | Sparse, discrete tokens embedded in graph-structured relational maps |
| **Core computation** | Predictive coding; interpolation over learned manifolds; precision-weighted inference | Symbol-based compositional inference |
| **Architecture** | Distributed, parallel, modality-specific modules; in-memory computation | Centralized, single-threaded; amodal convergence onto a common processing pathway |
| **Dynamics** | Continuous trajectories on the sensorimotor timescale; neural dynamics are the content | Punctate access events; dynamics serve as the vehicle |
| **Grounding & portability** | Tightly coupled to present sensory input; substrate-bound | Decoupled from the here-and-now; portable across brain regions, across time, and across individuals |
| **Scale properties** | Bounded by the manifold; scalar variability; smooth, magnitudinal errors | Scale-invariant; extrapolates indefinitely beyond training data; discrete, combinatorial failure modes |
| **Capacity** | High-bandwidth, massively parallel | Low-bandwidth, predominantly serial |
| **Neural substrates (candidate)** | Sensory and associative cortex across distributed cortical hierarchies | Anterior temporal lobe, medial temporal concept cells, entorhinal–hippocampal grid/graph machinery, frontoparietal control network, frontal pole |
| **Reportability** | Not directly reportable; contents remain subconscious unless translated | Directly reportable; contents available for verbal reports and symbolic manipulation |
| **Learning mechanism** | Gradual updating of network weights | Symbol crystallization and relational embedding via replay-driven consolidation |

**Table 1. Comparative summary of System C and System D across representational, computational, and functional dimensions.** The table consolidates the two-system distinction developed throughout the paper. System C denotes the continuous, high-dimensional substrate characterized by graded representations and predictive-coding dynamics (**Section 2.2**); System D denotes the discrete, categorical substrate (**Section 2.3**), characterized by symbol-like tokens and the geometric signature identified in **Section 4.1**. The contrasts should be read as idealized rather than sharp dichotomies, and the two systems are coupled via the C-to-D translation operation (**Section 3**). Neural substrate attributions are provisional and reflect current candidate mappings rather than settled commitments.

# 2. The C/D Framework: Redefining the Two-System Model

## 2.1 Why Fast/Slow Is Not Enough

The idea that the mind operates in two modes is among the most durable in cognitive science, spanning Wason and Evans's early dual-process proposals (Wason & Evans, 1974), Sloman's (1996) distinction between associative and rule-based reasoning, Stanovich and West's (2000) System 1/System 2 terminology, and Kahneman's (2011) popularization of "fast" and "slow" thinking. The framework has been productive, but its productivity has largely been in the domain of behavioral prediction. It tells us when people will be biased, when they will deliberate, when heuristics will dominate. What it does not tell us is what underlying neural mechanisms distinguish the two systems, or why the brain should be organized this way at all.

These limitations have been well recognized. Evans and Stanovich (2013) themselves conceded that "fast" and "slow" are misleading. For instance, System 2 processes are not always slow. Well-practiced logical inferences can be rapid. On the other hand, System 1 processes are not always fast. Perceptual and value-based decisions based on noisy sensory inputs can take seconds (Yang & Shadlen, 2007; Zhang et al., 2022). Melnikoff and Bargh (2018) went further, arguing that the defining features typically attributed to the two systems do not cluster as the theory requires. What looks like a natural kind at the level of phenomenology ("effortful deliberation" versus "intuitive response") dissolves into a heterogeneous collection of processes at the level of mechanism.

We propose that the genuine joint is representational. The brain is not divided into a fast system and a slow system; it is divided into a system that operates in continuous, high-dimensional manifolds and a system that runs on discrete, compositional symbols. Speed, effort, and phenomenological accessibility are downstream consequences of this representational difference, not its essence. Once the division is located at the representational level, features that seemed puzzling on the fast/slow framework fall into place as expected consequences of a single underlying architecture.

## 2.2 System C: The Continuous Engine

System C is the brain's default representational regime and its bedrock, shared in broad outline with almost every mobile organism with a nervous system. Its primary function is to map continuous sensory inputs to continuous motor outputs, enabling the organism to navigate and survive in a dynamic, high-dimensional physical environment. Its defining mechanistic feature is that information is carried in continuous variables, such as firing rates, synaptic weights, oscillatory phases, and that these variables are manipulated by operations like summation, multiplication, convolution, and gradual learning that respects the continuity of the underlying state space. In mathematical terms, System C operates over a continuous state space $X \subset \mathbb{R}^n$, with transformations from sensory input $s(t)$ to motor output $m(t)$ governed by continuous dynamics implemented in the synaptic weights of the neural network.

We should note that several components of this machinery are locally non-continuous. Action potentials are all-or-none events, synaptic release is quantal, and spike generation involves a threshold nonlinearity. We consider these discontinuities as biophysical implementation details rather than features of the representational format. What carries information is the graded population- and time-averaged quantities these events give rise to, which vary continuously with the stimuli and motor variables they encode.

### 2.2.1 Distributed Modularity and Evidence Accumulation

System C is not a single processor but a massive array of parallel, modality-specific modules distributed across cortex and subcortex. It does not possess a central executive, and it does not need one. The computational problems it solves are local, and the circuits that solve them are embedded in the specific sensory-motor loops from which their inputs arise.

Perceptual decision making provides a good example. There is no single "decision center" in System C; instead, decisions are embedded within the sensorimotor loop required to act on them. When an agent decides where to direct its gaze based on visual evidence, neurons in the lateral intraparietal area (LIP) and the frontal eye fields (FEF) accumulate sensory evidence over time, with firing rates ramping continuously until a threshold is reached and a saccade is triggered (Roitman & Shadlen, 2002; Gold & Shadlen, 2007; Kira et al., 2015). This dynamic can be described by drift-diffusion models (DDM), whose continuous decision variables directly mirror neural signals in these brain regions. When the same agent decides how to reach for an object with the hand, a different circuit entirely, including the parietal reach region (PRR) and dorsal premotor cortex, accumulates the spatial evidence relevant to that action (Andersen & Cui, 2009). Between the two cases, the computational signature is the same. Continuous quantities are integrated over time by circuits dedicated to the sensorimotor contingencies they serve.

The ubiquity of this organization across cortex (DiCarlo et al., 2012; Churchland et al., 2012; Stachenfeld et al., 2017; Bernardi et al., 2020) suggests that manifold-based distributed computation is a native format of neural computation. It also explains why System C is so well-suited to parallelism. It is exactly because its modules are physically distinct and operate on different continuous data streams. An organism can simultaneously accumulate evidence for visual tracking, adjust its posture via the cerebellum, and regulate its autonomic state via the brainstem, with little computational interference. Serial bottlenecks, when they appear in the cognitive architecture, are not a property of System C.

### 2.2.2 Interpolation Over Learned Manifolds

System C's computational signature is interpolation, and it excels at it. In machine learning and computational neuroscience alike, sensory inputs are understood to lie on low-dimensional continuous manifolds embedded within high-dimensional state spaces (DiCarlo & Cox, 2007). Interpolation generates predictions within the convex hull of training data on this manifold, and it is what such architectures do superbly well. An organism that has observed predators moving at 10 km/h and 20 km/h can smoothly infer the trajectory of a predator moving at 15 km/h. The computation is nearly free, because the manifold's local geometry contains the answer.

Within the envelope of experience the organism has actually sampled, System C provides fast, precise, graded predictions across a very wide range of situations, and the same continuous machinery that supports perception simultaneously supports action. When a human throws a ball, System C calculates the parabolic trajectory with effortless precision, integrating visual, proprioceptive, and vestibular signals into a coherent motor command in a few hundred milliseconds. No symbolic computation is required, and no symbolic computation would be as fast.

The regime of applicability of the interpolation is defined by where the manifold has been sampled. Outside of that regime, the same architecture is being asked to do something it was not trained to do. The question is therefore not whether System C is limited, but which class of problems calls for a representational format organized on different principles.

### 2.2.3 Grounding in the Sensorimotor Present

System C's representations are continuous and require continuous support. The manifold state encoding the visual layout of a room is maintained by the ongoing activity of the neurons that encode it, and when the sensory input is withdrawn that activity decays over a short period of time, sustained transiently by short-term synaptic facilitation and recurrent dynamics (Zylberberg & Strowbridge, 2017). This tight coupling between representation and ongoing sensory drive is not incidental. It is what allows System C to remain veridically bound to the present state of the world. A reactive engine optimized for immediate sensorimotor engagement should track the environment on the environment's own timescale, and System C does exactly that.

In System C, the neural dynamics themselves carry representational content. The temporal profile of a sensory signal is encoded in the temporal profile of the population trajectory that responds to it, and the trajectory of a motor output is directly driven by the corresponding trajectory of activity in premotor and motor circuits. Time-varying stimuli map to time-varying neural states, and time-varying neural states map to time-varying behavior, with the mapping preserved at the level of the dynamics rather than being abstracted away from them. This is why System C excels at tasks whose structure is inherently temporal, such as tracking a moving target, catching a ball, producing coordinated speech articulation, or executing a skilled reach. In these tasks, the "computation" is not separable from the trajectory through state space; the trajectory is the computation. The continuous format is uniquely suited to this regime because its state space is smooth, its updates respect that smoothness, and its outputs are naturally read out as continuous motor commands.

The representational temporality of System C is therefore the temporality of the environment it tracks, and this is what makes it so effective for the tasks it is asked to perform. Locomotion, postural control, gaze stabilization, reaching, and predator avoidance all require an engine that is continuously updated by current input and continuously coupled to current motor demands. Organisms whose fitness depends primarily on real-time interaction with a physical world receive from System C exactly what they need. Cognition that extends beyond the immediate sensorimotor envelope, such as reasoning about past episodes or future events on the scale of months and years, calls for a complementary representational format with different temporal properties.

### 2.2.4 In Memory Computation

System C, like neural networks generally, employs an in-memory computing architecture. There is no separation between where content is stored and where it is processed: the same synaptic weights that hold a learned regularity also perform the transformation that applies it, and the same population activity that represents a state is also the substrate on which the next computational step operates. This organization avoids the von Neumann bottleneck of shuttling data between memory and processor, permits massive parallelism, and is central to why biological neural computation achieves the speed and energy efficiency it does. For an engine dedicated to real-time sensorimotor control, in-memory computation is close to ideal: transformations are applied at the site of the representation, latencies are minimal, and every synapse contributes simultaneously to storage and to processing.

The cost of this architecture is that content encoded in it is inseparable from the substrate that carries it. A manifold state is defined only relative to the particular population of neurons and the particular connectivity that supports it, and stripped of that supporting context it is difficult to transfer to another system or another moment. This limitation manifests along three axes. Across the brain, two areas that both encode the same stimulus will nonetheless embed it in incongruent manifold geometries, so that coherent transmission depends on learned, pairwise-specific alignment between sender and receiver rather than on any general-purpose currency (Semedo et al., 2019). Across time, representational drift reshapes the population states supporting a given content over days to weeks even when behavior is stable (Driscoll et al., 2017; Rule et al., 2019; Schoonover et al., 2021), so that a downstream reader faces a moving target. Across individuals, two brains do not share a manifold coordinate system. The neurons that participate, the geometry into which content is embedded, and the projections by which it is read out are idiosyncratic to each brain, so a state copied directly from one brain into another would be lost in translation.

These three properties share a common structure. In each case the content of a System C representation is inseparable from the substrate that carries it and cannot be lifted off that substrate without loss. For an engine dedicated to real-time sensorimotor control the cost is negligible, since the relevant readers are immediately downstream circuits that have co-adapted with it and the relevant timescale is short enough that drift is inconsequential. It is not suitable for cognitive function that requires content to be routed flexibly across the brain, persisting stably across days, or shared between minds through language or teaching.

### 2.2.5 Summary

System C, then, is a distributed, parallel, continuously predictive engine that grounds the organism in its immediate environment through manifold-structured sensorimotor computation. Its representations are carried by high-dimensional population states whose trajectories directly encode the temporal profile of sensory input and directly drive the temporal profile of motor output, and its in-memory architecture co-locates storage and computation in the same synaptic and population substrate. Within its proper domain of perception, action, and the statistical regularities of the organism's experienced world, it is extraordinarily powerful, and, given unlimited resources, universally capable in principle. However, when extending cognition beyond these regimes, it calls for a complementary representational format organized on categorically different principles. That is the role we propose for System D.

## 2.3 System D: The Discrete Engine

### 2.3.1 Defining Symbols: Discrete, Scale-Invariant Categorical Identity

We define a symbol, for the purposes of this framework, as a discrete, scale-invariant categorical identity. A representation qualifies as symbolic when (i) it occupies one of a set of discrete categorical states rather than a position on a gradient, (ii) its computational role does not depend on the scale, intensity, or sensory modality of the input that produced it, and (iii) it can enter into compositional operations whose form is determined by syntactic structure, so that the same primitive can occupy multiple structural roles.

Crucially, "symbol" in this framework does not refer merely to linguistic tokens, such as words on a page or phonemes in speech. Those are external vehicles for symbols. The symbols are discrete categorical states implemented in specific neural populations. The linguistic token "dog" and the thought-constituent [DOG] are both symbols in our sense, but they are distinct instantiations of the same underlying representational type.

The discovery of concept cells in the human medial temporal lobe (Quiroga et al., 2005; Quiroga, 2012) provides neurobiological evidence. Recordings from single neurons in the hippocampus and entorhinal cortex have revealed cells that fire in a highly selective, invariant manner to abstract concepts. For example, a neuron responds to images of Jennifer Aniston, the written name "Jennifer Aniston," and the spoken name, but not to other faces, other names, or other voices. Importantly, the representation has been stripped of continuous sensory variation, and what remains is an amodal, discrete, categorical code. This sparse, invariant coding is exactly what the definition above requires, and it allows the brain to represent vast numbers of discrete concepts without the catastrophic interference that would result from overlapping continuous representations (Valdez et al., 2015).

Complementary evidence for the third criterion, which is the compositional operations governed by syntactic structure, comes from recent work on the Language of Problem Solving (LoPS) framework (Yang, Zhu, et al., 2025; Si, Yang, et al., 2026). Analyzing extended gameplay in an adapted Pac-Man paradigm, Yang and colleagues showed that both human and macaque problem-solving behavior can be parsed into a small, finite inventory of discrete primitive "strategies" that combine according to probabilistic grammar rules with well-defined temporal dependencies. Crucially, the primitives in this framework are not sensory categories or motor programs but latent policies defined over task structure. The same primitive can occupy different ordinal positions in a compositional sequence, and its functional role is determined by that syntactic position rather than by any continuous feature of the input. The complexity of an individual's inferred grammar predicted problem-solving performance and distinguished humans from macaques, whose grammars remained comparatively shallow, suggesting that the capacity to compose over abstract primitives is what scales with cognitive sophistication.

The follow-up fMRI study (Si, Yang, et al., 2026) localized these compositional operations to a right-lateralized frontoparietal circuit, with a nested hierarchy of representations in the brain. Grammar-rule transitions selectively engaged the right IFG and right SMG, the right-hemisphere homologues of Broca's and Wernicke's areas, and the functional coupling between these regions predicted individual differences in performance. The same architectural motif that supports linguistic syntax on the left appears to support non-linguistic compositional problem solving on the right.

For present purposes, this matters in two ways. First, it demonstrates that discrete, categorically identified units can enter into syntactic operations such as compositional bindings, thereby satisfying criterion (iii) with neurobiological rather than merely behavioral evidence. Second, it reinforces the notion that "symbol" is not a linguistic notion. The constituents composed by the right-lateralized LoPS circuit are not words or phonemes, yet they are discrete, position-sensitive, and compositionally combinable. Symbolic composition, on this evidence, is a domain-general capacity of which language is one lateralized specialization.

Converging single-unit evidence comes from Tian et al. (2026), who identified a population in macaque ventral premotor cortex whose activity during a drawing-like task exhibits categorical structure, invariance across effector and visual context, and recombination of the same primitive across varying sequential positions, satisfying our criteria of symbolic representation at the neuronal level. Together with the LoPS findings, this indicates that symbolic representation is neither confined to humans nor to the declarative domain, but is a general feature of frontal-cortical computation in primates.

This definition of symbol aligns in broad terms with the Language of Thought hypothesis (Fodor, 1975) and its recent revival by Quilty-Dunn and colleagues (2023), but it differs in a consequential respect. Where classical LOT treats symbols as the default currency of central cognition, we treat them as the product of a specific computational regime. Similarly, our definition is stronger than the notion of categorical perception in sensory cortex (Freedman et al., 2001), where category-sensitive neurons remain tied to a particular modality and scale. Symbolic representations in our sense have crossed a threshold that categorical sensory representations have not. They have severed their dependence on the continuous sensory substrate.

### 2.3.2 Spatiotemporal Decoupling and Mental Time Travel

The first and most consequential functional property of symbolic representation is spatiotemporal decoupling. Because a symbol is a discrete, compressed entity whose identity does not depend on ongoing sensory input, it can exist in the mind without being tethered to the present. Linguists have long recognized the capacity to refer to entities and events removed from the immediate perceptual field under the label displacement (Hockett, 1960). What linguistics describes at the level of external signs, we take to be a general property of the underlying symbolic format.

This decoupling has a specific mechanistic reading in light of the temporal-dynamics coding described in **Section 2.2.3**. In System C, the dynamics are the content. In System D, by contrast, the dynamics are the vehicle by which a token is ignited and held, not what the token represents. A concept cell's identity as [DOG] is not the shape of its firing trajectory but its position in a discrete relational scheme, and any dynamics that place the population into that discrete state realize the same representational content. Severing content from the moment-to-moment shape of neural dynamics is what allows a symbol to be decoupled from the here-and-now.

Spatiotemporal decoupling in this sense is what makes mental time travel possible (Tulving, 2002; Suddendorf & Corballis, 2007). System D can load symbolic tokens for "next year," "my childhood home," or "the trip into a black hole" into working memory and perform logical operations on them with no requirement that the referents be sensorily available. Episodic memory for specific past events, prospection about specific future events, and counterfactual reasoning about events that never occurred all exploit this decoupling.

Finally, spatiotemporal decoupling allows the separation of storage from computation. This is what makes System D content portable. Across the brain, a symbol can be evoked in one region and re-instantiated in another because both regions carry pointers into a shared discrete inventory rather than incongruent manifold geometries. Across time, a symbol's identity does not drift with the underlying population, because the identity is not defined by which specific neurons happen to constitute the assembly on a given day. Across individuals, two brains can communicate a symbol through language or teaching because the receiver need only bind an external vehicle to its own local realization of the same discrete role, without any requirement that the sender's and receiver's population geometries be aligned.

### 2.3.3 Scale Invariance and Infinite Extrapolation

The second defining property of symbolic representation is scale invariance. A symbol is scale-invariant in the sense that its computational role does not depend on the magnitude, duration, or granularity of the input from which it was extracted. The same token can participate in operations across scales that differ by many orders of magnitude, without retraining and without degradation.

The contrast is sharpest in a task both systems can attempt: simple addition. Consider $3.2 + 4.9$. System C can approximate this operation through the analog magnitude system that humans and many other animals share (Dehaene, 1997; Feigenson et al., 2004), and it will return an answer distributed around the correct value of $8.1$. The error is roughly Gaussian and its width is a smooth function of the operand magnitudes. Present the same system with $32000 + 49000$ and the distribution width grows with the signal, in accordance with the scalar variability that characterizes analog magnitude representations across species and modalities (Gallistel & Gelman, 2000). This is the interpolation regime, where the representation is tied to the scale of its inputs, and precision degrades as the inputs move away from the range in which the system was calibrated.

System D exhibits a qualitatively different error profile, and the difference is diagnostic of a different underlying format. When a person computing $3.2 + 4.9$ makes an error, the error is almost never a small continuous perturbation around $8.1$. It is a structured mistake. A subject who fails to propagate the carry returns $7.1$, and a subject who drops a decimal returns $81$. What no subject ever returns is $5.1$, or $11.4$, or any of the other values that a noisy analog computation could easily produce. The errors partition into a discrete set of failure modes, each traceable to the omission or misapplication of a specific symbolic sub-operation, because the computation itself is composed of discrete sub-operations over discrete tokens. When the scale changes, for instance, to $32000 + 49000$, the signature of each failure mode is preserved. The error is combinatorial rather than magnitudinal.

With symbols, System D extends to scales at which System C has no representation to interpolate over at all. System D can compute $3.2 \times 10^{1000} + 4.9 \times 10^{1000}$ and return $8.1 \times 10^{1000}$ with the same procedure, and the same error profile, as for $3.2 + 4.9$. System C, on the other hand, has no sensory experience of quantities of this magnitude, no manifold covering this range, and no analog approximation available. Symbolic representation allows the brain to extrapolate indefinitely beyond its biological training data, and to fail, when it fails, in ways that reveal the discrete structure of the operations being performed (Tenenbaum et al., 2011; Lake et al., 2017).

### 2.3.4 Compression, Convergence, and the Anterior Temporal Hub

The third property of symbolic representation is compression. A continuous manifold representation of a dog is the accumulated multi-modal high-dimensional structure of every interaction the organism has ever had with dogs. System D compresses this entire manifold neighborhood into a single lightweight token: [DOG]. The compression is many orders of magnitude and it is what allows working memory, with its severely limited capacity, to operate on concepts rather than on the sensory histories from which those concepts were distilled.

The neural implementation of this compression appears to be the hub-and-spoke architecture of semantic cognition (Patterson et al., 2007; Lambon Ralph et al., 2017). In this model, modality-specific sensory regions (the spokes) process continuous inputs in their respective formats, while a transmodal convergence zone, which is primarily the anterior temporal lobe (ATL), integrates across modalities to form representations in which the continuous sensory variation has been stripped away. The ATL is also the locus at which the dimensionality collapse characteristic of symbolic representation is implemented. Semantic dementia, which involves ATL atrophy, produces selective loss of conceptual identity with relative preservation of the underlying sensory manifolds, a dissociation the model predicts (Hodges & Patterson, 2007).

### 2.3.5 Representational Format: Sparse Codes and Relational Graphs

The three properties enumerated above, namely spatiotemporal decoupling, scale invariance, and compression, describe what System D representations do. Individual symbols are carried by sparse, combinatorial codes, and the symbols are embedded in graph-structured relational maps. Together these constitute the nodes and edges of the System D representational scheme.

The hub-and-spoke architecture just described, combined with the concept-cell findings from the medial temporal lobe, suggests that System D's individual representations are carried by a sparse, combinatorial coding scheme that stands in marked contrast to System C's distributed manifold coding. Where System C spreads representation across high-dimensional population activity, with each unit contributing in graded fashion to many representations, System D concentrates representation into small subsets of highly selective units drawn from a larger pool, with distinct symbols sharing units combinatorially.

A candidate computational formalism for how such codes are constructed and stabilized is the Assembly Calculus (Papadimitriou et al., 2020), in which discrete neural assemblies emerge from random recurrent connectivity through Hebbian plasticity and local inhibitory competition. Critically, the framework specifies a small set of operations on assemblies that are each reducible to biologically plausible spiking dynamics and provides, at the population-code level, a proposal for the format of System D representations.

To support inference and planning, discrete symbols must be embedded in a structure that specifies how they relate to one another. A neurobiological bridge has recently emerged from work on conceptual spaces. Grid cells in the entorhinal cortex, originally characterized for their role in mapping continuous physical space (Moser et al., 2008), exhibit the same hexagonal firing structure when subjects navigate abstract conceptual spaces, for instance a two-dimensional space of bird shapes varying in leg and neck length (Constantinescu et al., 2016; Bellmund et al., 2018). While these early demonstrations relied on continuously varying feature dimensions, subsequent work has extended the finding into genuinely discrete and graph-structured territory. The hippocampal-entorhinal system and ventromedial prefrontal cortex construct relational maps of discrete entities learned piecemeal across separate episodes (Park et al., 2020, 2021), and computational models formalize this circuitry as learning graph-structured transition rules over discrete latent states (Whittington et al., 2020). The implication is that System D repurposes the brain's spatial navigation hardware to structure relations among discrete concepts, traversing a graph of symbols with the same algorithms that previously guided traversal of physical terrain.

These two features function as complementary halves of a single representational format. Sparse combinatorial coding in the ATL and MTL supplies the nodes: discrete, categorically identified units carried by selective populations. The entorhinal-hippocampal graph machinery, together with its ventromedial prefrontal extensions, supplies the edges: a relational scaffold that positions these nodes with respect to one another in a structure over which transitive inference, planning, and analogy can be computed. The functional properties described in **Sections 2.3.2-2.3.4** are supported via this format. Spatiotemporal decoupling is available because sparse discrete codes do not require ongoing sensory drive to maintain their identity. Scale invariance follows because a node's role is defined by its position in the graph rather than by any continuous property of its input. Compression is what the format achieves in the first place, collapsing a high-dimensional manifold neighborhood into a small set of active units linked to others by relational edges. The difference between the two systems is therefore due to the computational primitives that the two formats make available.

### 2.3.6 Serial Access: The Single Thread

The picture developed so far raises an immediate puzzle about the two systems. A person can walk, listen to music, and visually scan the surroundings simultaneously and effortlessly (System C), yet the same person cannot consciously solve two simple mathematical problems in parallel (System D). The asymmetry is large, and it calls for explanation in terms of the architectural difference between the two formats.

We propose that while System C is distributed and parallel, System D is predominantly serial with a single-threaded, centralized architecture. By "centralized" we do not mean localized to a single brain region. The substrate may well be a distributed circuit spanning prefrontal, parietal, and subcortical structures. What matters is that the computation itself is carried out in a centralized manner. Symbolic content converges onto a common processing pathway that admits only one compositional operation at a time, regardless of how anatomically distributed its physical implementation may be. This serial bottleneck has been documented across paradigms under labels including the psychological refractory period (Pashler, 1994), the attentional blink (Raymond et al., 1992), and the capacity limits of working memory (Cowan, 2001).

We take this bottleneck to be a constitutive feature of symbolic operation in the human brain rather than an incidental limitation to be engineered away. Because System D must handle inputs from all modalities under a common logical structure, it cannot be distributed in the manner of System C. While System C can afford specialized parallel modules because each one operates on its own modality-specific manifold, System D cannot do so. A symbol that meant one thing in the visual stream and another in the auditory stream would fail the scale-invariance criterion by construction. Centralization is therefore a structural requirement of amodality in the brain. And once the architecture is centralized, parallel processing of distinct logical streams through the same hub produces the serial bottleneck.

#### 2.3.7 Summary

System D, then, is a discrete, symbolic, centralized, and serial representational regime whose neurobiological substrate includes the anterior temporal lobe (as a transmodal compression hub), the medial temporal lobe (as a concept-cell archive), the entorhinal grid system (as the scaffolding for structured conceptual traversal), and the frontoparietal control network together with the frontal pole (as the execution machinery for sequential symbolic search). Its computational virtues, including scale invariance, compositional generativity, spatiotemporal decoupling, portable and addressable content, and extrapolation and generalization, complement what System C excels at.

# 3. Conscious Access as Representational Translation

## 3.1 The Mismatch Problem

The architecture developed in **Section 2** leaves us with a specific computational problem. The high-dimensional continuous manifolds used in System C and the discrete, scale-invariant, amodal symbols used in System D are almost maximally dissimilar. Yet the cognitive capacities that distinguish human intelligence, including reportable perception, deliberate reasoning, planning, language, and counterfactual thought, depend on information transfer between them.

We propose that conscious access requires the computational operation that performs this transfer. More specifically, we hypothesize that conscious access is marked by the structure-preserving map from a localized neighborhood of a System C manifold to a discrete token in System D, together with the inverse projection from System D tokens back onto System C manifolds. It is neither a property of System C in isolation nor of System D in isolation; it is the translation between them. When we say a mental state is "consciously accessed," we mean that the state in question has been rendered in both formats simultaneously and is available for report. The continuous manifold activity and the discrete symbolic token are, at that moment, placed into structural correspondence.

This framing specifies what the neural correlates of conscious access should look like. It is not a specific brain region or a specific oscillation, but a specific representational geometry in which continuous manifold structure and discrete categorical structure are placed into correspondence.

Consequently, conscious access should be measurable from the neural dynamics accompanying the translation.

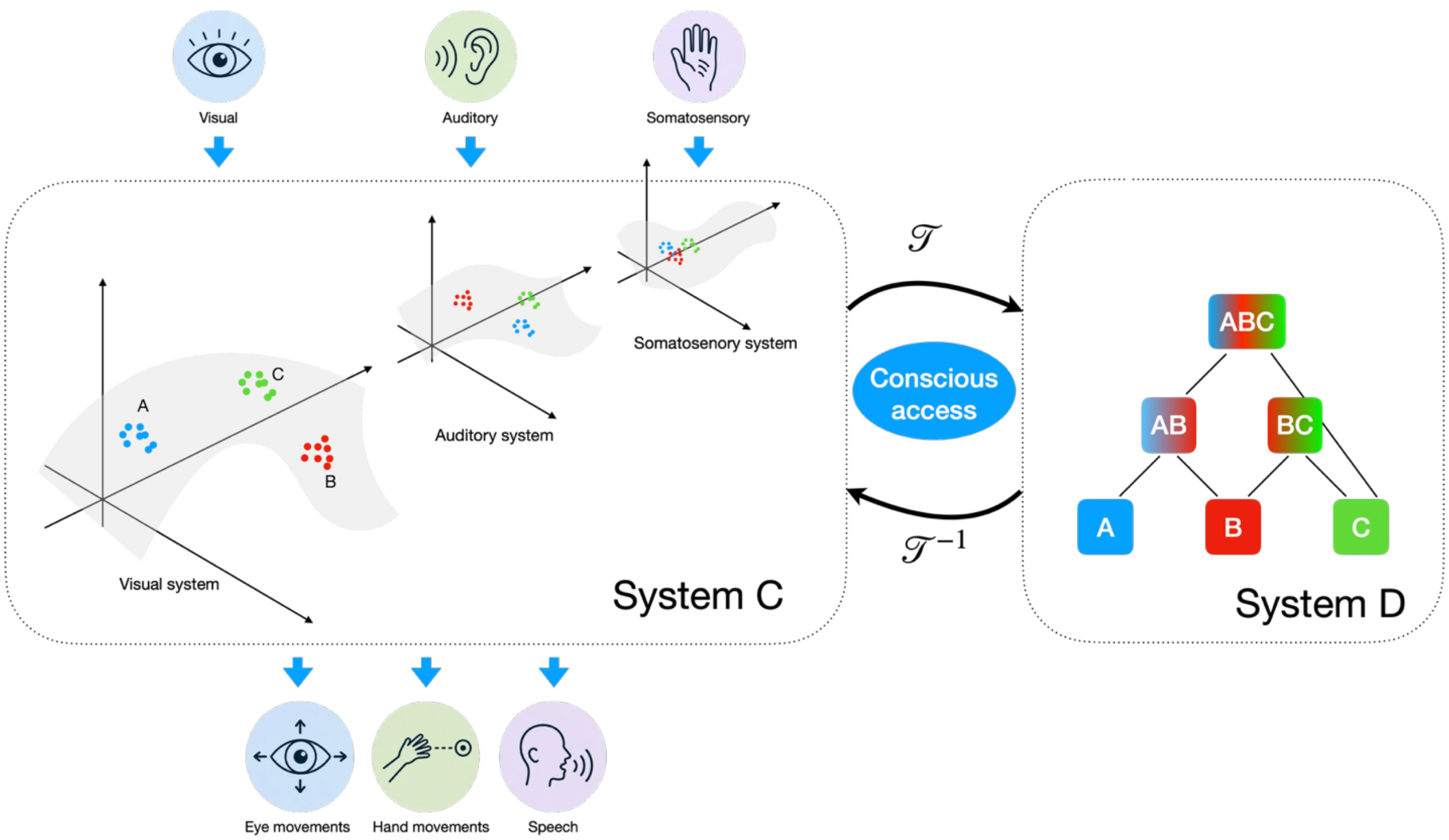

**Figure 1. Schematic of the C/D framework.** The figure depicts the two representational substrates posited by the framework and the bidirectional translation operation that couples them. Sensory inputs from multiple modalities drive activity within System C, a continuous, high-dimensional substrate distributed across modality-specific subspaces in parallel (left panels). Within each subspace, recurring stimulus categories give rise to clustered activity patterns (colored point clouds labeled A, B, C) embedded in a graded manifold geometry. Cross-modal correspondences are indicated by consistent color coding across the visual, auditory, and somatosensory panels. The operation $\mathcal{T}$ implements categorical collapse, mapping continuous manifold states onto discrete tokens in System D (right). We identify this operation with conscious access. System D is organized as a compositional structure over discrete tokens (A, B, C), supporting compositional (AB, BC, ABC) groupings that inherit the categorical identity of their constituents. The inverse operation $\mathcal{T}^{-1}$ re-embeds discrete tokens back into the manifold, allowing System D content to shape ongoing C-dynamics. Downstream effectors (eye movements, hand movements, speech) are driven from System C and modulated by the System D content re-embedded via $\mathcal{T}^{-1}$.

## 3.2 What the Translation Operation Is

Formally, let $\mathcal{M} \subset \mathbb{R}^n$ denote a manifold of System C states and let $\mathcal{S}$ denote the (effectively) discrete set of symbolic tokens available in System D. The C-to-D translation is a map

$$\mathcal{T}: \mathcal{N}(x) \subset \mathcal{M} \longrightarrow s \in \mathcal{S}$$

that assigns, to a neighborhood $\mathcal{N}(x)$ around a point $x$ on the manifold, a single symbolic token $s$. The map is many-to-one by construction. An entire neighborhood of continuous states collapses onto a single discrete token. This compression is what allows downstream symbolic operations to run over a small inventory of stable units rather than over the full continuous state space, and it is the mechanism by which category identity is made invariant to within-category variation.

Two features of $\mathcal{T}$ are worth emphasizing. First, $\mathcal{T}$ is structure-preserving in the sense relevant to downstream computation. Tokens whose corresponding manifold neighborhoods are close in the continuous space will, under normal conditions, stand in correspondingly close relations in the symbolic graph, though the metric of the symbolic space is not simply inherited from the manifold. The representational geometry of the symbolic layer is a coarse-grained, topologically simplified version of the representational geometry of the underlying manifold.

The projection operation of the Assembly Calculus (Papadimitriou et al., 2020) is a candidate model of $\mathcal{T}$. The input pattern of firing in System C drives, via random feedforward projections and recurrent Hebbian dynamics, the convergence onto a sparse stable assembly that is deterministic given the input and discrete in the sense that distinct inputs yield either the same assembly or a well-separated one. These are precisely the properties we require of $\mathcal{T}$.

Second, $\mathcal{T}$ admits an "inverse" projection

$$\mathcal{T}^{-1}: s \in \mathcal{S} \longrightarrow \mathcal{N}'(x') \subset \mathcal{M}$$

where $\mathcal{N}'(x')$ denotes a re-instantiated neighborhood, not necessarily identical to the original $\mathcal{N}(x)$. Note here the term inverse is used loosely, as $\mathcal{T}$ is many-to-one and compressive. Its fine-grained continuous detail of any particular manifold state is not recoverable from the symbol alone and must be supplied by the priors of System C and by whatever contextual information is available.

This inverse projection, we propose, is the computational identity of mental imagery and simulated perception. When a subject "imagines" an apple, System D tokens are projected back onto System C manifolds in roughly the regions that sensory perception of an apple would activate, but with the fine-grained continuous detail of real perception partially hallucinated by System C's generative machinery rather than supplied by sensory input (Pearson, 2019; Dijkstra et al., 2019).

## 3.3 Neural Implementation

We have so far characterized the translation operation abstractly. The question of where and how it is implemented in the brain is empirical and, at present, incompletely resolved. We sketch here the circuit-level hypothesis the framework naturally suggests, noting that aspects of it remain speculative. The proposals are offered as falsifiable hypotheses.

Four mechanistic questions organize the sketch. If System D is a single-threaded bottleneck, what determines which continuous representations are granted entry? How are discrete symbols constructed in the first place from the continuous experiential history that precedes them? Once constructed, how are individual symbols organized into the relational structure that supports compositional inference? And by what dynamical principles does a continuous manifold state collapse, on any given occasion, into a discrete symbolic node? Our answers invoke, respectively, the basal ganglia–thalamic gating circuit, the hippocampal–neocortical dialogue of systems consolidation, replay-driven construction of graph-structured cognitive maps, and the attractor-like dynamics that produce sparse, discrete assemblies from continuous input.

### 3.3.1 Gating

A single-threaded architecture embedded in a massively parallel substrate requires a gate that decides, on a millisecond timescale, which of the many active manifold states in System C enters the narrow channel of System D. The thalamic reticular nucleus (TRN) is well positioned for this role, since virtually all thalamocortical traffic passes under its inhibitory influence (Pinault, 2004; Halassa & Kastner, 2017), and its selective suppression and disinhibition of thalamic relays modulates which continuous sensory streams reach cortical targets with high fidelity. The PBWM (prefrontal cortex basal ganglia working memory) framework (Frank et al., 2001; O'Reilly & Frank, 2006) provides a well worked-out account. The striatum evaluates the motivational relevance of incoming representations, weighted by dopaminergic reward prediction errors, and when a representation is sufficiently salient, disinhibits the thalamus via the globus pallidus internus, allowing the selected representation to occupy one of the limited slots of System D. Striatal evaluation is continuous and parallel, but disinhibition is discrete and its downstream consequence is categorical. Therefore, the basal ganglia-TRN circuit performs, at the level of gating, the same continuous-to-discrete transition that characterizes access more generally. This predicts that basal ganglia dysfunction should produce characteristic deficits in the selection of contents for access without necessarily disrupting the translation operation once selection has occurred.

### 3.3.2 Symbol construction

Gating explains how existing symbols are selected; it does not explain how symbols are constructed. An infant encountering dogs does not have the concept [DOG] pre-installed. It is distilled over time from continuous episodic experiences. The Complementary Learning Systems framework provides the scaffolding (McClelland et al., 1995; Kumaran et al., 2016). The hippocampus supports rapid one-shot encoding of continuous episodic traces, while the neocortex supports slow, interleaved learning of statistical regularities extracted across many such episodes.

As formulated, however, CLS is primarily a theory of System C learning: it describes how continuous statistical regularities are extracted and stabilized in transmodal neocortex. It does not, on its own, explain how such regularities become the discrete symbols that System D operates on. If symbol construction were driven purely by sensory statistics, the brain would be overwhelmed by every minor variation in texture, lighting, or background noise present in the training manifold. Instead, we propose that symbol formation is driven by a demand for compositionality in problem solving and decision making. A representation is not "symbolized" simply because it is frequent. It becomes a discrete symbol only if it satisfies two criteria. First, the representation must correspond to distinct outcomes, affordances, or causal powers in the environment. Second, the representation must possess structural properties that make it useful as an operand in compositional reasoning.

The construction of new symbols is therefore not a passive process but shaped by downstream computational needs. As the hippocampus replays episodic sequences to the neocortex during offline consolidation (Buzsáki, 2015), transmodal neocortical circuitry such as the ATL does not merely average out sensory noise. It extracts invariants that are "worth" retaining because they maximize the efficiency of System D operations. The ATL thus serves as a selective crystallization site where continuous history is pruned to retain only those features that bridge effectively between perception and action (Patterson et al., 2007; Lambon Ralph et al., 2017). This explains why we construct symbols for foods or predators but do not spontaneously generate symbols for random fluctuations in background sensory noise, which has low utility or compositional value. The inventory of System D is therefore dynamic: new nodes may form, and the translation function $\mathcal{T}$ must correspondingly extend to cover them.

### 3.3.3 Relational embedding

Individual symbols, however well-formed, are not yet a system. For System D to support compositional reasoning, symbols must be embedded in a relational structure consisting of transitions, similarities, and causal dependencies over which inferences can be computed. Consolidation supplies this structure. Offline consolidation operates largely through hippocampal sharp-wave ripples during slow-wave sleep and quiet rest, replaying recently encoded sequences to the neocortex in temporally compressed form (Buzsáki, 2015).

Crucially, replay is not confined to spatial trajectories. The human hippocampus sequentially replays non-spatial task states during rest (Schuck & Niv, 2019), and replayed sequences are spontaneously reorganized according to learned relational structure rather than merely recapitulating the order of experience (Liu et al., 2019). Replay thus functions as a general mechanism for learning graph-structured cognitive maps over discrete states (Momennejad et al., 2018; Mattar & Daw, 2018). This process is what gives System D's symbolic inventory its shape as a graph. Symbols crystallized by the mechanisms discussed above are, through replay-driven consolidation, positioned relative to one another in a structure that supports transitive inference, planning, and analogical generalization. Construction supplies the nodes; consolidation supplies the edges.

The two processes together predict a specific dissociation. Disruptions to consolidation should selectively impair both the construction of new symbols and the relational reorganization of existing ones, while leaving translation intact for previously consolidated concepts and previously learned relations. This matches the pattern observed in hippocampal amnesia, where established semantic knowledge and its relational structure are preserved but new acquisition is severely impaired (Gabrieli et al., 1988; Manns et al., 2003).

### 3.3.4 Categorical commitment

Granting that symbols exist as stable patterns in neocortical circuitry, what is the mechanism by which a continuous input is converted into a discrete output? One candidate mechanisms is the projection operation of the Assembly Calculus (Papadimitriou et al., 2020). Assembly dynamics can be viewed as a specific, biologically constrained form of attractor dynamics (Hopfield, 1982; Wang, 2002; Khona & Fiete 2022). A pattern of firing in an upstream area drives, via random feedforward connections and recurrent dynamics in a downstream area, the convergence onto a sparse stable assembly whose identity is deterministic given the input, invariant under small perturbations, and discretely separated from the assemblies evoked by categorically distinct inputs. The commitment is therefore not a phase transition imposed on top of the representation but is intrinsic to the dynamics that construct it. It can serve as a promising model for the neurobiological basis of categorical perception (Harnad, 1987; Freedman et al., 2001) and a general consequence of the translation operation applied to continuous input.

Therefore, the commitment mechanism and the representational format may not be two separate things. A symbol is the sparse attractor state, and the same dynamics that construct it also make it composable. Candidate cortical substrates include the anterior temporal lobe as an amodal semantic hub (Patterson et al., 2007), medial temporal structures supporting invariant concept-like representations (Quiroga, 2012), and lateral prefrontal cortex for task-relevant categorical decisions (Freedman et al., 2001), with the specific target of projection determined by the feedforward connectivity of the source area.

## 3.4 Summary

Conscious access, on the C/D framework, is the computational operation that translates localized neighborhoods of System C manifolds into discrete System D tokens, together with the inverse projection from tokens back onto manifolds. The operation is structure-preserving at the level of similarity geometry, lossy at the level of fine-grained continuous detail, and implemented by a distributed circuit in which thalamocortical loops gate entry, hippocampal-neocortical consolidation shapes the categorical landscape, and sparse combinatorial codes in transmodal temporal cortex carry the resulting symbols. It is the operation that GNWT, IIT, PP, and HOT each partially describe without fully isolating.

# 4. Empirical Predictions and Comparison with Existing Frameworks

## 4.1 The Core Prediction: A Geometric Signature of Translation

The C/D framework makes a single unifying empirical prediction from which all others are derived. We propose that conscious access is marked by a specific transformation in representational geometry, identifiable as a collapse from high-dimensional continuous similarity structure to low-dimensional, near-categorical similarity structure, measured at the transition between System C and System D. For any set of stimuli, this transformation should be detectable in the geometry, topology, and dynamics of population activity in the regions bridging sensory cortex and symbolic substrates. Formally, if the population state in System C substrates lies on a continuous manifold reflecting graded physical properties, and the population state in System D substrates uses a sparse, distributed combinatorial code, then the translation operation $\mathcal{T}$ should be recoverable as the nonlinear compression that best predicts the second geometry from the first, preserving categorical structure while discarding within-category continuous variation.

This single prediction yields a family of more specific hypotheses:

**H1** (Format change). Successful conscious access will be associated with a significant reduction in the intrinsic dimensionality of neural representation as one moves from sensory to symbolic substrates, measured as the dimensionality of the respective representational spaces. This reduction should be substantially larger for consciously accessed stimuli than for matched unreported stimuli.

**H2** (Categorical collapse). Within-category dissimilarity in System D substrates should fall below within-category dissimilarity in System C substrates for consciously accessed stimuli, while between-category dissimilarity should be preserved or enhanced. This yields a specific prediction about the within- to between-category dissimilarity across the translation.

**H3** (Reportability coupling). The degree of categorical collapse measured in **H2** should predict trial-by-trial reportability. On trials in which the subject reports a given stimulus, the collapse should be pronounced; on trials in which the subject fails to report an otherwise identical stimulus, the collapse should be attenuated or absent. This is a within-subject, within-stimulus prediction that can be tested in paradigms with matched physical input and varying report (inattentional blindness, attentional blink, binocular rivalry, masking, etc.).

**H4** (Temporal ordering). The categorical collapse in System D substrates should lag the corresponding continuous representation in System C substrates by a measurable interval, corresponding to the duration of the translation operation. This interval should be compressible under practice but should not vanish. It also sets a lower bound on the latency of conscious access.

**H5** (Inverse projection). When subjects perform mental imagery or working-memory maintenance of a specific stimulus, the representational geometry in System C substrates should partially recapitulate the sensory-evoked geometry, but with reduced dimensionality, lower fidelity and preserved categorical structure, which are signatures of the inverse projection $\mathcal{T}^{-1}$.

This is certainly not an exhaustive list. Other predictions follow from the same core commitment, for instance, concerning the computational cost of translation, the effects of disruption of D-substrate integrity, cross-species comparisons, and learning-induced changes in representational geometry. We foreground H1–H5 because they are the most directly testable with existing methods and collectively cover several important aspects of the translation operation, but they are best understood as a starting point.

Testing these hypotheses requires a battery of complementary analyses, each sensitive to a distinct facet of the translation. Representational similarity analysis (Kriegeskorte et al., 2008; Kriegeskorte & Kievit, 2013) indexes the geometric shift addressed by **H2** by comparing empirical RDMs to model RDMs corresponding to graded physical similarity versus discrete block structure, and may help recover the transformation $\mathcal{T}$ relating C- and D-substrate geometries. Cross-condition generalization performance (CCGP; Bernardi et al., 2020), together with shattering dimensionality, distinguishes symbolic from merely compressed codes by locating a representation in a two-dimensional geometry space. C-format codes should be high-shattering and low-CCGP, D-format codes the reverse, providing a joint test of **H1** and **H2**. Intrinsic dimensionality estimators (participation ratio, TwoNN; Facco et al., 2017) provide a scalar summary of the compression predicted by **H1** that can be tracked efficiently across regions and time. Persistent homology (Chaudhuri et al., 2019) targets the qualitative topological change predicted in **Section 3.3**. Temporal generalization analysis (King & Dehaene, 2014) tests the dynamical prediction underlying **H4** by revealing whether C-format states produce transient diagonal decoding while D-format states produce sustained block-structured generalization. These methods can be applied on the same data, cross-referenced against the reportability contrast in **H3** and the imagery contrast in **H5**. None of these five hypotheses is uniquely predicted by GNWT, IIT, PP, or HOT, and several of them are directly inconsistent with the strongest versions of each.

## 4.2 Contrast with Global Neuronal Workspace Theory

GNWT, in its mature formulation (Dehaene et al., 2017; Mashour et al., 2020), holds that conscious access occurs when a sensory representation is "ignited" in frontoparietal cortex and broadcast globally through long-range connections. The signature empirical marker is a late (~300 ms), non-linear amplification of stimulus-specific activity in frontoparietal regions, accompanied by widespread distributed activation and, in several studies, a characteristic "P3b" event-related potential.

The C/D framework shares substantial empirical ground with GNWT. Both predict frontoparietal involvement, a late ignition-like event distinguishing accessed from unaccessed stimuli, and flexible cross-brain routing of accessed content. The frameworks diverge on what the decisive event consists in. For GNWT, the decisive event is broadcast, which is the propagation of a representation from local sensory circuits into the global workspace. For C/D, the decisive event is the translation of a continuous manifold state into a discrete symbolic token. Broadcast, on the other hand, is a downstream consequence of successful translation. Once a representation has been created symbolically it can be routed anywhere, but broadcasting a state that has not been translated does not constitute conscious access.

This difference generates three discriminating predictions, each targeting a distinct component of the conjoint signature outlined in **Section 4.1**. First, broadcast alone should be insufficient for access. Frontoparietal ignition of a representation whose underlying code remains continuous and high-dimensional should fail to support reportability, discrete manipulation, and integration with symbolic working memory. This is a direct test of **H1** and **H2**. Second, the critical representational transition should be observed in temporal cortex. RSA and CCGP applied to Cogitate-style paradigms (Cogitate Consortium et al., 2025) should reveal the categorical collapse signature (**H2**) arising earlier and more sharply in anterior and medial temporal regions, with persistent homology detecting the topological change in these regions. Third, P3b-like ignition should dissociate from conscious access when translation proceeds via automatized routes. Highly practiced categorization (e.g., native-language word recognition) may accomplish the C-to-D translation with minimal late frontoparietal engagement, in which case the geometric and dynamical signatures (**H1**, **H2**, **H4**) should be preserved even as the P3b is attenuated. This can be measured via CCGP, dimensionality reduction, and temporal generalization and is an outcome that GNWT has difficulty accommodating.

## 4.3 Contrast with Integrated Information Theory

IIT (Oizumi et al., 2014; Tononi et al., 2016) holds that consciousness is identical to integrated information (Φ), which is  a measure of the degree to which a system generates more information as a whole than as the sum of its parts.

Both the C/D and IIT frameworks agree that conscious states have a distinctive representational structure, and both agree that this structure is measurable in principle. Where they diverge is that IIT locates consciousness in the intrinsic causal structure of the substrate, independent of representational format, whereas C/D locates conscious access in the translation between two specific formats.

The two frameworks generate opposed empirical predictions on two fronts. First, integration and translation should dissociate. Systems that achieve high Φ within a continuous format but do not implement the C-to-D translation should not exhibit the behavioral profile of conscious access, including geometric compression (**H1**), categorical collapse (**H2**), or temporal dynamics (**H4**). In such systems, intrinsic dimensionality should remain high, CCGP should remain low, and temporal generalization should show transient rather than sustained regimes even where Φ is maximized. Second, the representational geometry at the moment of access should exhibit categorical discretization, not smooth high-Φ integration. IIT predicts increased representational richness in regions implementing the maximal complex, whereas C/D predicts a compression signature. These predictions concern different features of the same data and can be evaluated jointly.

## 4.4 Contrast with Predictive Processing

PP, in its consciousness-relevant formulations (Hohwy, 2013; Clark, 2019; Seth, 2021), holds that conscious experience is constituted by the brain's best-guess hierarchical model of the causes of its sensory input, with precision-weighting of prediction errors modulating the content and vividness of experience. On the strongest versions, conscious experience is the generative model in operation.

The relationship between the C/D framework and PP is more cooperative than competitive. We endorse predictive coding as an account of how System C operates. Our disagreement is with the extension of PP into a theory of conscious access. Continuous prediction is not sufficient for conscious access. What the PP does not by iteslf specify is the format change.

This yields a discriminating prediction. Precision modulation within System C that does not cross the C-to-D translation should not produce the geometric signature (**H2**) characteristic of access, even as it alters sensory representations. According to PP, increases in precision-weighting should directly amplify a representation's contribution to experience, since experience is constituted by the precision-weighted generative model itself. The C/D framework predicts that precision changes within System C modulate the input to the translation operation but do not themselves constitute access, and a high-precision continuous representation that fails to trigger the categorical collapse in System D substrates should remain non-reportable and unavailable for symbolic manipulation. This prediction is testable in paradigms that manipulate attention or expectation to alter precision-weighting while independently measuring representational geometry at both System C and System D substrates.

More broadly, we take PP and the C/D framework to be operating at different levels of the same architecture rather than competing for the same explanatory territory. PP characterizes the computations within System C. The C/D framework characterizes the operation that renders System C representations available to System D. A complete theory will need both.

## 4.5 Contrast with Higher-Order Theories

The three frameworks considered so far offer points of contrast with C/D. The higher-order tradition, by contrast, offers a point of kinship. The core HOT claim is that a mental state becomes conscious in virtue of being represented by a suitable higher-order state, whether characterized as a higher-order thought (Rosenthal, 2005), a self-model (Fleming, 2020), or a perceptual reality monitor (Lau & Rosenthal, 2011; Brown, Lau, & LeDoux, 2019). The higher-order state is typically localized to prefrontal and adjacent transmodal cortex, and its disruption is invoked to explain dissociations between first-order representation and conscious report in blindsight, inattentional blindness, and no-report paradigms. Thus, unlike the other major theories, HOT and C/D both propose that conscious access involves a re-representation of first-order content in a distinct neural and functional locus. The C/D framework specifies a higher-order state is associated with conscious access only when it supports symbolic abstraction, categorical commitment, and bidirectional grounding. This is what classical HOT has left open and departs from it on specific theoretical commitments while sharing its central insight.

The most outstanding departure concerns whether metarepresentation is by itself sufficient for conscious access. HOT, on its constitutive reading, holds that it is. The empirical picture invites a more restricted reading. Implicit metacognition in humans dissociates from conscious report, with confidence signals, error-related activity, and neural signatures of decision uncertainty present in conditions where subjects show no explicit metacognitive access (Charles et al., 2013; Fleming & Dolan, 2012). Non-human animals show robust metacognitive behavior on uncertainty-monitoring tasks, opting out of trials when perceptual evidence is weak and wagering less on low-confidence responses, and the underlying confidence signals have been identified in parietal and orbitofrontal circuits that support perceptual and value-based decision-making rather than in the prefrontal higher-order machinery HOT typically invokes (Kiani & Shadlen, 2009; Lak et al., 2014; Vivar-Lazo & Fetsch, 2026). Modern artificial neural networks produce confidence estimates over their own outputs, and many architectures include explicit metacognitive components, without any plausible higher-order phenomenology. The computational structure of metarepresentation can therefore be implemented entirely within continuous manifold dynamics, in circuits that are not plausibly the seat of conscious access and without any transition into a distinct representational format.

These findings do not undermine HOT's central insight, but they do suggest a dissociation between metarepresentation and conscious access. The C/D framework identifies the missing specification as the format transition itself. What determines conscious access is not any metarepresentation but the translation into a categorical, discrete, sparsely coded, compositional format.

The two frameworks therefore make overlapping but non-identical empirical predictions. Both predict that disruption of the higher-order or translation circuit should selectively impair report and metacognition while sparing first-order representation. Where they diverge is in the format signature of the surviving first-order state. The C/D framework predicts that unreported states should retain the high-dimensional continuous similarity geometry that categorical collapse would otherwise flatten. The C/D framework also predicts that there is no conscious access when metacognition shows continuous-format signatures, such as graded confidence, and absence of categorical collapse, while metacognition accessible to consciousness should show the categorical signature characteristic of System D more generally. In contrast, HOT leaves the format of the higher-order state largely unspecified and characterizes the higher-order relation in terms that are not directly related to neural activity.

### 4.6 Summary

The C/D framework makes a unified, geometrically specified empirical prediction. Conscious access is the moment at which a continuous representational manifold is translated into a discrete symbolic code, and this moment has distinctive signatures, including dimensionality reduction, categorical collapse, topological simplification, and dynamical stabilization. The framework shares ground with GNWT on the involvement of frontoparietal cortex, with IIT on the representational specificity of conscious states, with PP on the importance of predictive inference in sensorimotor processing, and with HOT on the re-representational structure of conscious access. It disagrees with each on what the decisive event of conscious access consists in. It is not broadcast, integration, precision-weighting, or metarepresentation alone, but fundamentally a change of representational format. These disagreements generate testable predictions that can, in several cases, be evaluated on existing data and, in others, are within reach of future experimental paradigms.

## 5. Implications for Artificial General Intelligence

### 5.1 Two Traditions

The history of artificial intelligence has been, to a large degree, the history of two research traditions. The classical symbolic tradition, ranging from early production systems through expert systems and formal knowledge representation, built machinery that operated on discrete, compositional tokens with well-defined syntactic properties (Newell & Simon, 1976; Russell & Norvig, 2020). These systems manipulated symbols with precision and supported genuine compositional generalization, but their symbols were defined purely by their syntactic relations to other symbols. They could not extract new categories from raw sensory input, and they possessed no principled way to connect their internal representations to the continuous physical world. The connectionist tradition, including neural networks, deep learning, and their contemporary descendants, on the other hand, built machinery that operated on continuous high-dimensional representations learned from data, and these systems proved extraordinarily effective at perception, pattern recognition, and within-distribution generalization (LeCun et al., 2015). But they were equally limited in a dual sense. They failed to support robust compositional generalization, they extrapolated poorly beyond their training distribution, and their internal representations, while rich, were not naturally decomposable into discrete reusable units.

The two traditions also diverged on how their representations were acquired. Classical symbolic systems typically treated symbols as pre-given by the designer and did not learn them from experience at all. Connectionist systems learned their continuous representations from data, but the learning procedure was offline and over a fixed corpus, with no mechanism for the ongoing construction of discrete units from new experience. Neither tradition implemented the pattern that the C/D framework identifies as central: a continuous substrate that updates on the timescale of experience and a discrete substrate whose inventory grows through consolidation, with a translation between the two that extends as the inventory grows.

The C/D framework offers a diagnosis and a solution. Each tradition built something like a single System C or D without the translation between them and was incomplete about the architecture as a whole. The neuro-symbolic research program that has emerged over the past decade (Garcez et al., 2019; Marcus, 2020; Mao et al., 2019; Yang, Shao et al., 2025) correctly identifies the need for both formats, but it has largely treated their integration as a matter of interface design, getting symbolic and neural modules to communicate. The framework developed in this paper suggests that the translation is not an interface problem but the central problem, and that the specific properties we have identified in biological implementation are what make the integration work.

## 5.2 Diagnosing Large Language Models

The most prominent contemporary AI systems are the large language models (LLMs) and their multimodal successors. These systems have demonstrated capabilities that would have seemed implausible a decade ago. They have also exhibited characteristic failures that the C/D framework can diagnose.

LLMs are, architecturally, vast continuous systems trained to predict tokens in a discrete linguistic stream, and they occupy a distinctive position relative to the C/D architecture. Their inputs and outputs are discrete tokens, but their internal operations run over continuous embeddings, and the tokens they manipulate are defined by their distributional relations to other tokens rather than by grounding in sensorimotor manifolds or by embedding in a sparse, amodal symbolic inventory. In C/D terms, an LLM is a very large System C whose training signal has been the output of human System D operations, filtered through language. What it has learned to do, with remarkable success, is to reproduce the surface statistics of symbolic behavior using continuous machinery.

This architectural profile predicts both the successes and the failures of these systems. Their successes lie where the distributional patterns of human symbolic behavior are rich and consistent, for instance, fluent generation, plausible completion of common reasoning patterns, and translation between languages. The tasks on which they are less reliable are those in which surface distributional patterns diverge from the discrete structure of the underlying operations, including arithmetic with out-of-distribution digit lengths (Nogueira et al., 2021), compositional generalization to novel combinations of familiar components (Lake & Baroni, 2018), and any task requiring a stable discrete state to persist across steps of reasoning. These are the tasks for which the continuous format alone is not the natural regime of applicability, and they are exactly the tasks that, in biological cognition, recruit the scale-invariant compositional machinery of System D.

The architectural diagnosis extends to how these systems learn. Biological cognition, as characterized in **Section 3.3**, learns continuously. The hippocampus encodes new episodes on the timescale of experience, consolidation reshapes the discrete symbol inventory over hours to years, and the translation function $\mathcal{T}$ is updated as new symbols crystallize and old ones are relationally re-embedded. An LLM has no analogue of this ongoing update, because it has no analogue of the two substrates with the correspondence $\mathcal{T}$. Pretraining is a massive offline optimization over a fixed corpus, after which the weights are frozen for deployment. During inference, there is no update to the model at all. What looks like learning is in-context learning, a temporary reweighting of attention over tokens supplied in the prompt, discarded when the context window closes. Fine-tuning and reinforcement learning from human feedback can extend the training phase intermittently, but they preserve the fundamental separation between an offline period in which the model changes and an online period in which it does not.

Prompt-based methods including chain-of-thought (Wei et al., 2022) and retrieval augmentation partially compensate by offloading fast-learning demands onto external text. This is an effective engineering strategy, and it accounts for much of the recent improvement in multi-step reasoning benchmarks. But it is architecturally inverted from the biological arrangement, where fast and slow learning are both internal to the system and coupled through consolidation. The scratchpad is outside the model, in a format the model must re-parse through the same frozen weights each time it is consulted, and nothing that appears on it durably modifies the system that reads it.

On the C/D framework, then, current systems have implemented, with remarkable engineering success, a version of System C sufficient to reproduce the outputs of symbolic cognition without implementing the underlying architecture and without the continuous learning regime that architecture requires. The regime in which this succeeds is broader than most observers would have anticipated a decade ago. The regime in which it does not is equally informative, and its boundary maps onto the class of problems for which the continuous format is not, on our account, the natural substrate (Marcus, 2020; Mitchell, 2021; Mitchell & Krakauer, 2023).

## 5.3 Architectural Prescriptions

Here, we propose a compact set of architectural commitments for systems aspiring beyond current LLMs based on the C/D framework. An adequate architecture must implement two representational formats as constitutively distinct components: a continuous manifold-based substrate (System C) and a discrete sparse-code substrate (System D). The properties that make System D computationally valuable (scale invariance, compositionality, resistance to catastrophic interference) are properties of the discrete format itself and cannot be approximated by a continuous network producing discrete-looking outputs. The two substrates must be connected by a bidirectional, structure-preserving translation operation implementing both the forward projection $\mathcal{T}$ and its inverse $\mathcal{T}^{-1}$, learned and dynamically updating rather than fixed.

These commitments distinguish the C/D framework sharply from the existing landscape of neuro-symbolic AI. Contemporary neuro-symbolic systems can be roughly grouped into three families, and each engages a different aspect of the integration problem. Reading them through the C/D framework helps clarify which aspect of the problem each addresses and which remains open.

The first family comprises cascaded architectures, in which a neural front-end performs perception and a downstream symbolic module performs reasoning over its outputs (Kautz, 2022; Sheth et al., 2023). These systems treat the two substrates as separable modules coupled at their interface, rather than as two formats of a single representational system linked by a learned translation. The symbolic vocabulary is typically fixed in advance rather than extracted from continuous experience, and information flow is predominantly unidirectional. The second family comprises differentiable-logic approaches such as DeepProbLog, Logic Tensor Networks, and Neural Theorem Provers, which embed symbolic operations inside a continuous computational graph so that logical structure can be trained end-to-end. These systems are technically impressive but at the cost of blurring the very format distinction we take to be essential. Symbols become soft vectors in a continuous space, and the categorical, scale-invariant properties that make System D valuable are only approximated. The third family comprises tool-augmented and code-generating LLMs, in which a continuous generative model produces discrete outputs that are then executed externally. Here the external layer does not learn during operation, and is unable to feed back into the continuous substrate in a structure-preserving way.

The three families together cover a substantial part of the design space, but the specific configuration C/D identifies as central is absent. None of them implements two constitutively distinct formats, coupled by a bidirectional and online-learned translation whose symbolic inventory is extracted from experience. Closed-loop systems that lack such coupling to a stable symbolic ground have been shown to exhibit characteristic degenerative dynamics (Zenil, 2026), and unified symbol grounding across explicit and implicit representations remains the central open problem.

Finally, the properties enumerated above cannot be established at design time and then frozen. They should be sustained by continuous online learning across both substrates and across the translation itself. System C must update on the timescale of experience, adjusting the geometry of its continuous manifolds as new sensorimotor data arrive. System D must support both rapid one-shot binding of new episodic content and slow consolidation-driven crystallization of new symbols, with $\mathcal{T}$ extending to cover the inventory as it grows and $\mathcal{T}^{-1}$ correspondingly extended. The training-deployment distinction, which dominates current large-scale AI, must be substantially softened. After an initial acquisition phase, the system must continue to construct symbols, revise relational structure, and update the translation operation throughout its operational lifetime.

Taken together, these commitments describe an architecture that differs in configuration from both current LLMs and current neuro-symbolic hybrids. What is central is the specific relation between the two substrates that differ constitutively in representational format, coupled by a structure-preserving operation. This operation runs in both directions, extracts its symbolic vocabulary from experience rather than inheriting it, and updates online as the system's experience accumulates.

# 6. Conclusion

The mind exhibits two apparently incompatible faces: continuous, high-dimensional, parallel processing of the kind that supports perception and motor control, and discrete, compositional, serial processing of the kind that supports language and deliberative reasoning. We have argued that these are two constitutively distinct substrates connected by a specific computational operation, and that we may model this translation as conscious access. The forward projection $\mathcal{T}$ compresses continuous manifold neighborhoods into discrete tokens through extraction of statistical invariants; the inverse projection $\mathcal{T}^{-1}$ grounds those tokens in simulated sensorimotor activity. The operation is lossy, categorical, and bidirectional, implemented by a coordinated circuit spanning basal ganglia-thalamic gating, hippocampal-neocortical consolidation, and anterior temporal sparse, distributed representation, with empirically testable predictions.

We do not claim to have solved the problems of consciousness, symbol grounding, or general intelligence. We claim that they share an architectural structure, that this structure can be specified with useful precision, and that doing so yields concrete traction on each. The continuous and the discrete are not competing theories of mind. They are two halves of a single architecture, and the operation that connects them is where much of what we call thinking actually happens.

# Acknowledgments

We thank Wangzhou Dai, Qianli Yang, and Qing Yu for helpful discussions on the ideas in this paper, and Yusi Chen and Fenrong Liu for a critical reading of the manuscript. This work was supported by the National Science and Technology Innovation 2030 Major Program (grant no. 2021ZD0203701) and the Strategic Priority Research Program of the Chinese Academy of Sciences (grant no. XDB1010301).

# References

Andersen, R. A., & Cui, H. (2009). Intention, action planning, and decision making in parietal-frontal circuits. *Neuron*, *63*(5), 568–583.

Bellmund, J. L. S., Gärdenfors, P., Moser, E. I., & Doeller, C. F. (2018). Navigating cognition: Spatial codes for human thinking. *Science*, *362*(6415).

Bernardi, S., Benna, M. K., Rigotti, M., Munuera, J., Fusi, S., & Salzman, C. D. (2020). The geometry of abstraction in the hippocampus and prefrontal cortex. *Cell*, *183*(4), 954–967.

Brown, R., Lau, H., & LeDoux, J. E. (2019). Understanding the higher-order approach to consciousness. *Trends in Cognitive Sciences*, *23*(9), 754–768.

Buzsáki, G. (2015). Hippocampal sharp wave-ripple: A cognitive biomarker for episodic memory and planning. *Hippocampus*, *25*(10), 1073–1188.

Chalmers, D. J. (1995). Facing up to the problem of consciousness. *Journal of Consciousness Studies*, *2*(3), 200–219.

Charles, L., Van Opstal, F., Marti, S., & Dehaene, S. (2013). Distinct brain mechanisms for conscious versus subliminal error detection. *NeuroImage*, *73*, 80–94.

Chaudhuri, R., Gerçek, B., Pandey, B., Peyrache, A., & Fiete, I. (2019). The intrinsic attractor manifold and population dynamics of a canonical cognitive circuit across waking and sleep. *Nature Neuroscience*, *22*(9), 1512–1520.

Churchland, M. M., Cunningham, J. P., Kaufman, M. T., Foster, J. D., Nuyujukian, P., Ryu, S. I., & Shenoy, K. V. (2012). Neural population dynamics during reaching. *Nature*, *487*(7405), 51–56.

Clark, A. (2019). Consciousness as generative entanglement. *Journal of Philosophy*, *116*(12), 645–662.

Cogitate Consortium, Ferrante, O., Gorska-Klimowska, U., … Melloni, L. (2025). Adversarial testing of global neuronal workspace and integrated information theories of consciousness. *Nature*, *642*(8066), 133–142.

Constantinescu, A. O., O'Reilly, J. X., & Behrens, T. E. J. (2016). Organizing conceptual knowledge in humans with a gridlike code. *Science*, *352*(6292), 1464–1468.

Cowan, N. (2001). The magical number 4 in short-term memory: A reconsideration of mental storage capacity. *Behavioral and Brain Sciences*, *24*(1), 87–114.

Dehaene, S. (1997). *The number sense: How the mind creates mathematics*. Oxford University Press.

Dehaene, S., Lau, H., & Kouider, S. (2017). What is consciousness, and could machines have it? *Science*, *358*(6362), 486–492.

Dehaene, S., & Naccache, L. (2001). Towards a cognitive neuroscience of consciousness: Basic evidence and a workspace framework. *Cognition*, *79*(1–2), 1–37.

DiCarlo, J. J., & Cox, D. D. (2007). Untangling invariant object recognition. *Trends in Cognitive Sciences*, *11*(8), 333–341.

DiCarlo, J. J., Zoccolan, D., & Rust, N. C. (2012). How does the brain solve visual object recognition? *Neuron*, *73*(3), 415–434.

Dijkstra, N., Bosch, S. E., & van Gerven, M. A. J. (2019). Shared neural mechanisms of visual perception and imagery. *Trends in Cognitive Sciences*, *23*(5), 423–434.

Driscoll, L. N., Pettit, N. L., Minderer, M., Chettih, S. N., & Harvey, C. D. (2017). Dynamic reorganization of neuronal activity patterns in parietal cortex. *Cell*, *170*(5), 986–999.

Evans, J. St. B. T., & Stanovich, K. E. (2013). Dual-process theories of higher cognition: Advancing the debate. *Perspectives on Psychological Science*, *8*(3), 223–241.

Facco, E., d'Errico, M., Rodriguez, A., & Laio, A. (2017). Estimating the intrinsic dimension of datasets by a minimal neighborhood information. *Scientific Reports*, *7*, 12140.

Feigenson, L., Dehaene, S., & Spelke, E. (2004). Core systems of number. *Trends in Cognitive Sciences*, 8(7), 307–314.

Fleming, S. M. (2020). Awareness as inference in a higher-order state space. *Neuroscience of Consciousness*, *2020*(1).

Fleming, S. M., & Dolan, R. J. (2012). The neural basis of metacognitive ability. *Philosophical Transactions of the Royal Society B*, *367*(1594), 1338–1349.

Fodor, J. A. (1975). *The language of thought*. Harvard University Press.

Frank, M. J., Loughry, B., & O'Reilly, R. C. (2001). Interactions between frontal cortex and basal ganglia in working memory: A computational model. *Cognitive, Affective, & Behavioral Neuroscience*, *1*(2), 137–160.

Freedman, D. J., Riesenhuber, M., Poggio, T., & Miller, E. K. (2001). Categorical representation of visual stimuli in the primate prefrontal cortex. *Science*, *291*(5502), 312–316.

Gabrieli, J. D. E., Cohen, N. J., & Corkin, S. (1988). The impaired learning of semantic knowledge following bilateral medial temporal-lobe resection. *Brain and Cognition*, *7*(2), 157–177.

Gallistel, C. R., & Gelman, R. (2000). Non-verbal numerical cognition: From reals to integers. *Trends in Cognitive Sciences*, 4(2), 59–65.

Garcez, A. d'A., Gori, M., Lamb, L. C., Serafini, L., Spranger, M., & Tran, S. N. (2019). Neural-symbolic computing: An effective methodology for principled integration of machine learning and reasoning. *Journal of Applied Logics*, *6*(4), 611–632.

Gold, J. I., & Shadlen, M. N. (2007). The neural basis of decision making. *Annual Review of Neuroscience*, *30*, 535–574.

Halassa, M. M., & Kastner, S. (2017). Thalamic functions in distributed cognitive control. *Nature Neuroscience*, *20*(12), 1669–1679.

Harnad, S. (1987). *Categorical perception: The groundwork of cognition*. Cambridge University Press.

Hockett, C. F. (1960). The origin of speech. *Scientific American*, *203*(3), 88–96.

Hodges, J. R., & Patterson, K. (2007). Semantic dementia: A unique clinicopathological syndrome. *The Lancet Neurology*, *6*(11), 1004–1014.

Hohwy, J. (2013). *The predictive mind*. Oxford University Press.

Hopfield, J. J. (1982). Neural networks and physical systems with emergent collective computational abilities. *Proceedings of the National Academy of Sciences*, *79*(8), 2554–2558.

Kahneman, D. (2011). *Thinking, fast and slow*. Farrar, Straus and Giroux.

Kautz, H. (2022). The third AI summer: AAAI Robert S. Engelmore memorial lecture. *AI Magazine*, 43(1), 105–125.

Khona, M., & Fiete, I. R. (2022). Attractor and integrator networks in the brain. *Nature Reviews Neuroscience*, *23*(12), 744–766.

Kiani, R., & Shadlen, M. N. (2009). Representation of confidence associated with a decision by neurons in the parietal cortex. *Science*, *324*(5928), 759–764.

King, J.-R., & Dehaene, S. (2014). Characterizing the dynamics of mental representations: The temporal generalization method. *Trends in Cognitive Sciences*, *18*(4), 203–210.

Kira, S., Yang, T., & Shadlen, M. N. (2015). A neural implementation of Wald's sequential probability ratio test. *Neuron*, *85*(4), 861–873.

Kriegeskorte, N., & Kievit, R. A. (2013). Representational geometry: Integrating cognition, computation, and the brain. *Trends in Cognitive Sciences*, *17*(8), 401–412.

Kriegeskorte, N., Mur, M., & Bandettini, P. (2008). Representational similarity analysis — connecting the branches of systems neuroscience. *Frontiers in Systems Neuroscience*, *2*, 4.

Kumaran, D., Hassabis, D., & McClelland, J. L. (2016). What learning systems do intelligent agents need? Complementary learning systems theory updated. *Trends in Cognitive Sciences*, *20*(7), 512–534.

Lak, A., Costa, G. M., Romberg, E., Koulakov, A. A., Mainen, Z. F., & Kepecs, A. (2014). Orbitofrontal cortex is required for optimal waiting based on decision confidence. *Neuron*, *84*(1), 190–201.

Lake, B. M., & Baroni, M. (2018). Generalization without systematicity: On the compositional skills of sequence-to-sequence recurrent networks. *Proceedings of the 35th International Conference on Machine Learning (ICML)*, 2873–2882.

Lake, B. M., Ullman, T. D., Tenenbaum, J. B., & Gershman, S. J. (2017). Building machines that learn and think like people. *Behavioral and Brain Sciences*, *40*, e253.

Lambon Ralph, M. A., Jefferies, E., Patterson, K., & Rogers, T. T. (2017). The neural and computational bases of semantic cognition. *Nature Reviews Neuroscience*, *18*(1), 42–55.

Lau, H., & Rosenthal, D. (2011). Empirical support for higher-order theories of conscious awareness. *Trends in Cognitive Sciences*, *15*(8), 365–373.

LeCun, Y., Bengio, Y., & Hinton, G. (2015). Deep learning. *Nature*, *521*(7553), 436–444.

Liu, Y., Dolan, R. J., Kurth-Nelson, Z., & Behrens, T. E. J. (2019). Human replay spontaneously reorganizes experience. *Cell*, *178*(3), 640–652.

Manns, J. R., Hopkins, R. O., Reed, J. M., Kitchener, E. G., & Squire, L. R. (2003). Recognition memory and the human hippocampus. *Neuron*, *37*(1), 171–180.

Mao, J., Gan, C., Kohli, P., Tenenbaum, J. B., & Wu, J. (2019). The neuro-symbolic concept learner: Interpreting scenes, words, and sentences from natural supervision. *Proceedings of the International Conference on Learning Representations (ICLR)*.

Marcus, G. (2020). The next decade in AI: Four steps towards robust artificial intelligence. *arXiv*:2002.06177.

Mashour, G. A., Roelfsema, P., Changeux, J.-P., & Dehaene, S. (2020). Conscious processing and the global neuronal workspace hypothesis. *Neuron*, *105*(5), 776–798.

Mattar, M. G., & Daw, N. D. (2018). Prioritized memory access explains planning and hippocampal replay. *Nature Neuroscience*, *21*(11), 1609–1617.

McClelland, J. L., McNaughton, B. L., & O'Reilly, R. C. (1995). Why there are complementary learning systems in the hippocampus and neocortex: Insights from the successes and failures of connectionist models of learning and memory. *Psychological Review*, *102*(3), 419–457.

Melnikoff, D. E., & Bargh, J. A. (2018). The mythical number two. *Trends in Cognitive Sciences*, *22*(4), 280–293.

Misak, C. (2013). *The American pragmatists*. Oxford University Press.

Mitchell, M. (2021). Why AI is harder than we think. *arXiv*:2104.12871.

Mitchell, M., & Krakauer, D. C. (2023). The debate over understanding in AI's large language models. *Proceedings of the National Academy of Sciences*, *120*(13).

Momennejad, I., Otto, A. R., Daw, N. D., & Norman, K. A. (2018). Offline replay supports planning in human reinforcement learning. *eLife*, 7, e32548.

Moser, E. I., Kropff, E., & Moser, M.-B. (2008). Place cells, grid cells, and the brain's spatial representation system. *Annual Review of Neuroscience*, *31*, 69–89.

Newell, A., & Simon, H. A. (1976). Computer science as empirical inquiry: Symbols and search. *Communications of the ACM*, *19*(3), 113–126.

Nogueira, R., Jiang, Z., & Lin, J. (2021). Investigating the limitations of transformers with simple arithmetic tasks. *arXiv*:2102.13019.

Oizumi, M., Albantakis, L., & Tononi, G. (2014). From the phenomenology to the mechanisms of consciousness: Integrated Information Theory 3.0. *PLOS Computational Biology*, *10*(5), e1003588.

O'Reilly, R. C., & Frank, M. J. (2006). Making working memory work: A computational model of learning in the prefrontal cortex and basal ganglia. *Neural Computation*, *18*(2), 283–328.

Papadimitriou, C. H., Vempala, S. S., Mitropolsky, D., Collins, M., & Maass, W. (2020). Brain computation by assemblies of neurons. *Proceedings of the National Academy of Sciences*, *117*(25), 14464–14472.

Park, S. A., Miller, D. S., Nili, H., Ranganath, C., & Boorman, E. D. (2020). Map making: Constructing, combining, and inferring on abstract cognitive maps. *Neuron*, *107*(6), 1226–1238.

Park, S. A., Miller, D. S., & Boorman, E. D. (2021). Inferences on a multidimensional social hierarchy use a grid-like code. *Nature Neuroscience*, *24*(9), 1292–1301.

Pashler, H. (1994). Dual-task interference in simple tasks: Data and theory. *Psychological Bulletin*, *116*(2), 220–244.

Patterson, K., Nestor, P. J., & Rogers, T. T. (2007). Where do you know what you know? The representation of semantic knowledge in the human brain. *Nature Reviews Neuroscience*, *8*(12), 976–987.

Pearson, J. (2019). The human imagination: The cognitive neuroscience of visual mental imagery. *Nature Reviews Neuroscience*, *20*(10), 624–634.

Pinault, D. (2004). The thalamic reticular nucleus: Structure, function and concept. *Brain Research Reviews*, *46*(1), 1–31.

Putnam, H. (1995). *Pragmatism: An open question*. Blackwell.

Quilty-Dunn, J., Porot, N., & Mandelbaum, E. (2023). The best game in town: The reemergence of the language-of-thought hypothesis across the cognitive sciences. *Behavioral and Brain Sciences*, *46*, e261.

Quiroga, R. Q. (2012). Concept cells: The building blocks of declarative memory functions. *Nature Reviews Neuroscience*, *13*(8), 587–597.

Quiroga, R. Q., Reddy, L., Kreiman, G., Koch, C., & Fried, I. (2005). Invariant visual representation by single neurons in the human brain. *Nature*, *435*(7045), 1102–1107.

Raymond, J. E., Shapiro, K. L., & Arnell, K. M. (1992). Temporary suppression of visual processing in an RSVP task: An attentional blink? *Journal of Experimental Psychology: Human Perception and Performance*, *18*(3), 849–860.

Roitman, J. D., & Shadlen, M. N. (2002). Response of neurons in the lateral intraparietal area during a combined visual discrimination reaction time task. *Journal of Neuroscience*, *22*(21), 9475–9489.

Rosenthal, D. M. (2005). *Consciousness and Mind*. Oxford: Clarendon Press.

Rule, M. E., O'Leary, T., & Harvey, C. D. (2019). Causes and consequences of representational drift. *Current Opinion in Neurobiology*, 58, 141–147.

Russell, S. J., & Norvig, P. (2020). *Artificial intelligence: A modern approach* (4th ed.). Pearson.

Schoonover, C. E., Ohashi, S. N., Axel, R., & Fink, A. J. P. (2021). Representational drift in primary olfactory cortex. *Nature*, 594(7864), 541–546.

Schuck, N. W., & Niv, Y. (2019). Sequential replay of nonspatial task states in the human hippocampus. *Science*, *364*(6447).

Semedo, J. D., Zandvakili, A., Machens, C. K., Yu, B. M., & Kohn, A. (2019). Cortical areas interact through a communication subspace. *Neuron*, *102*(1), 249–259.

Seth, A. K. (2021). *Being you: A new science of consciousness*. Dutton.

Sheth, A., Roy, K., & Gaur, M. (2023). Neurosymbolic AI — Why, what, and how. *IEEE Intelligent Systems*, 38(3), 56–62. *arXiv*:2305.00813.

Si, R., Yang, Q., Zhu, Z., Zhang, J., & Yang, T. (2026). Structured problem-solving recruits language-like hierarchical brain network. *SSRN*, [http://dx.doi.org/10.2139/ssrn.6036986](http://dx.doi.org/10.2139/ssrn.6036986)

Sloman, S. A. (1996). The empirical case for two systems of reasoning. *Psychological Bulletin*, *119*(1), 3–22.

Stachenfeld, K. L., Botvinick, M. M., & Gershman, S. J. (2017). The hippocampus as a predictive map. *Nature Neuroscience*, *20*(11), 1643–1653.

Stanovich, K. E., & West, R. F. (2000). Individual differences in reasoning: Implications for the rationality debate? *Behavioral and Brain Sciences*, *23*(5), 645–665.

Suddendorf, T., & Corballis, M. C. (2007). The evolution of foresight: What is mental time travel, and is it unique to humans? *Behavioral and Brain Sciences*, *30*(3), 299–313.

Tenenbaum, J. B., Kemp, C., Griffiths, T. L., & Goodman, N. D. (2011). How to grow a mind: Statistics, structure, and abstraction. *Science*, *331*(6022), 1279–1285.

Tian, L. Y., Garzón Gupta, K., Hanuska, D. J., Rouse, A. G., Eldridge, M. A. G., Schieber, M. H., Wang, X. J., Tenenbaum, J. B., & Freiwald, W. A. (2026). Neural representation of action symbols in primate frontal cortex. *Nature*, *654*(8117).

Tononi, G., Boly, M., Massimini, M., & Koch, C. (2016). Integrated information theory: From consciousness to its physical substrate. *Nature Reviews Neuroscience*, *17*(7), 450–461.

Tulving, E. (2002). Episodic memory: From mind to brain. *Annual Review of Psychology*, *53*(1), 1–25.

Valdez, A. B., Papesh, M. H., Treiman, D. M., Smith, K. A., Goldinger, S. D., & Steinmetz, P. N. (2015). Distributed representation of visual objects by single neurons in the human brain. *Journal of Neuroscience*, *35*(13), 5180–5186.

Vivar-Lazo, M., & Fetsch, C. R. (2026). Neural basis of concurrent deliberation toward a choice and confidence judgment. *Nature Neuroscience*, *29*(1),159-170.

Wang, X.-J. (2002). Probabilistic decision making by slow reverberation in cortical circuits. *Neuron*, *36*(5), 955–968.

Wason, P. C., & Evans, J. St. B. T. (1974). Dual processes in reasoning? *Cognition*, *3*(2), 141–154.

Wei, J., Wang, X., Schuurmans, D., Bosma, M., Ichter, B., Xia, F., Chi, E., Le, Q., & Zhou, D. (2022). Chain-of-thought prompting elicits reasoning in large language models. *Advances in Neural Information Processing Systems*, *35*, 24824–24837.

Whittington, J. C. R., Muller, T. H., Mark, S., Chen, G., Barry, C., Burgess, N., & Behrens, T. E. J. (2020). The Tolman-Eichenbaum Machine: Unifying space and relational memory through generalization in the hippocampal formation. *Cell*, *183*(5), 1249–1263.

Yang, Q., Zhu, Z., Si, R., Li, Y., Zhang, J., & Yang, T. (2025). A language model of problem solving in humans and macaque monkeys. *Current Biology*, *35*(1), 11–20.e10.

Yang, T., & Shadlen, M. N. (2007). Probabilistic reasoning by neurons. *Nature*, *447*(7148), 1075–1080.

Yang, X. W., Shao, J. J., Guo, L. Z., Zhang, B. W., Zhou, Z., Jia, L. H., ... & Li, Y. F. (2025). Neuro-symbolic artificial intelligence: Towards improving the reasoning abilities of large language models. *arXiv*:2508.13678.

Zhang, Z., Yin, C., & Yang, T. (2022). Evidence accumulation occurs locally in the parietal cortex. *Nature Communications*, *13*(1), 4426.

Zenil, H. (2026). On the limits of self-improving in large language models: The singularity is not near without symbolic model synthesis. *arXiv*:2601.05280v2.

Zylberberg, J., & Strowbridge, B. W. (2017). Mechanisms of persistent activity in cortical circuits: Possible neural substrates for working memory. *Annual Review of Neuroscience*, *40*, 603–627.